\documentclass[
  10pt,
  longbibliography,
  twocolumn,
  twoside,
]{article}

\usepackage[numbers,round,comma,sort&compress]{natbib}

\usepackage[
  left=0.65in,
  right=0.65in,
  top=0.65in,
  bottom=0.65in,
]{geometry}

\usepackage{sectsty}
\sectionfont{\raggedright\large\bfseries}
\subsectionfont{\raggedright\large}

\usepackage[font=small,labelfont=bf]{caption}

\usepackage{fancyhdr}
\usepackage[realmainfile]{currfile}
\input{\currfilebase.settings}
\usepackage{amssymb}
\usepackage{amsmath}

\usepackage{url}
\PassOptionsToPackage{hyphens}{url}\usepackage{hyperref}
\makeatletter
\g@addto@macro{\UrlBreaks}{\UrlOrds}
\makeatother

\usepackage{colortbl}

\definecolor{goodblue}{RGB}{0, 91, 187}
\usepackage{hyperref}
\hypersetup{
  colorlinks=true,
  allcolors=goodblue,
  urlcolor=goodblue,
  citecolor=goodblue,
  pdfborder={0 0 0},
  breaklinks=true,
}

\usepackage[normalem]{ulem}

\usepackage{textcomp}

\usepackage[export]{adjustbox}

\makeatletter

\def\CT@@do@color{%
  \global\let\CT@do@color\relax
  \@tempdima\wd\z@
  \advance\@tempdima\@tempdimb
  \advance\@tempdima\@tempdimc
  \advance\@tempdimb\tabcolsep
  \advance\@tempdimc\tabcolsep
  \advance\@tempdima2\tabcolsep
  \kern-\@tempdimb
  \leaders\vrule
  \hskip\@tempdima\@plus  1fill
  \kern-\@tempdimc
  \hskip-\wd\z@ \@plus -1fill }
\makeatother

\usepackage[table]{xcolor}

\definecolor{olivegreen}{rgb}{0.33333,.41961,0.18431}
\definecolor{forestgreen}{rgb}{0.13333,.5451,0.13333}
\definecolor{lightgrey}{rgb}{0.7,0.7,0.7}
\definecolor{verylightgrey}{rgb}{0.90,0.90,0.90}
\definecolor{veryverylightgrey}{rgb}{0.95,0.95,0.95}
\definecolor{grey}{rgb}{0.5,0.5,0.5}

\definecolor{headerblue}{HTML}{33367E}
\definecolor{unitednationsblue}{HTML}{4D88FF}

\definecolor{charcoal}{HTML}{36454F}
\definecolor{cinerous}{HTML}{98817B}
\definecolor{feldgrau}{HTML}{4D5D53}
\definecolor{glaucous}{HTML}{6082B6}
\definecolor{arsenic}{HTML}{3B444B}
\definecolor{xanadu}{HTML}{738678}

\definecolor{firebrick}{HTML}{B22222}
\definecolor{orangered}{HTML}{FF4500}
\definecolor{tomato}{HTML}{FF6347}

\definecolor{purpletaupe}{HTML}{3B444B}

\newcommand{\done}[1]{}

\newenvironment{textblock}{\renewcommand{\item}{}}{}

\usepackage{titlesec}
\titleformat*{\paragraph}{\bfseries}

\usepackage{changepage}

\usepackage{graphicx}
\usepackage{epsfig}
\usepackage{verbatim}

\usepackage{enumerate}
\usepackage{enumitem}

\usepackage{amssymb}
\usepackage{amsmath}

\usepackage{ifthen}

\usepackage{longtable}

\usepackage{mathtools}

\usepackage{tikz}
\usetikzlibrary{shapes.geometric, arrows}
\tikzstyle{startstop} = [rectangle, rounded corners, minimum width=4cm, minimum height=1cm,text centered, text centered, text width=5cm, draw=black, fill=red!30]
\tikzstyle{arrow} = [thick,->,>=stealth]

\newboolean{twocolswitch}

\usepackage{array}

\newcommand{\PreserveBackslash}[1]{\let\temp=\\#1\let\\=\temp}

\newcommand{\sindex}[1]{}
\newcommand{\nindex}[1]{}

\newcommand{\www}[1]{\url{#1}}

\usepackage{lettrine}

\newcommand{\lmat}{\left[
    \begin{array}
    }
    \newcommand{\rmat}{\end{array}
  \right]
}

\setboolean{twocolswitch}{true}

\usepackage{natbib}
\setcitestyle{square}
\raggedright

\usepackage{authblk}

\begin{document}

\title{\protect
Hollywood's misrepresentation of death: 
A comparison of overall and by-gender 
mortality causes in film and the real world







}


\renewcommand*{\Authsep}{, }
\renewcommand*{\Authand}{, }
\renewcommand*{\Authands}{, }
\renewcommand*{\Affilfont}{\normalsize\normalfont}
\renewcommand*{\Authfont}{}
\setlength{\affilsep}{2em}

\author[1,2]{Calla~Beauregard}
\author[1,2,3]{Christopher~M.~Danforth}
\author[1,2,4,5]{Peter~Sheridan~Dodds}

\affil[1]{
  Computational Story Lab,
  Vermont Advanced Computing Center,
  University of Vermont,
  Burlington,
  VT 05405,
  US
}

\affil[2]{
  Vermont Complex Systems Center,
  MassMutual Center of Excellence for Complex Systems and Data Science,
  University of Vermont,
  Burlington,
  VT 05405,
  US
  }

\affil[3]{
  Department of Mathematics \& Statistics,
  University of Vermont,
  Burlington,
  VT 05405,
  US
  }

\affil[4]{
  Department of Computer Science,
  University of Vermont,
  Burlington,
  VT 05405,
  US
}

\affil[5]{
  Santa Fe Institute,
  1399 Hyde Park Rd,
  Santa Fe,
  NM 87501,
  US
}

\date{\today}

\maketitle



\mbox{}

\bigskip
\bigskip
\bigskip

\hspace*{+1.5in}
\begin{minipage}{4in}

  \begin{tabular}{>{\raggedright\arraybackslash}p{180pt}p{180pt}}
    \renewcommand{\baselinestretch}{1.25}
    \selectfont
    
    \textbf{Logline}
    
    \bigskip

    For the top 10 major causes of death in the United States, we examine how cinematic representation of overall and by-gender mortality diverges from reality. Movies strongly overrepresent suicide and, to a lesser degree, accidents. By gender, we find that media largely overrepresents male mortality and underrepresents female mortality, particularly for heart disease and cerebrovascular disease; the two exceptions for which women are overrepresented are
suicide and accidents. 

    \renewcommand{\baselinestretch}{1}
    \selectfont
    
  \end{tabular}
  
\end{minipage}

\clearpage

\newgeometry{
  left=2in,
  right=2in,
  top=1in,
  bottom=1in,
  }

\onecolumn

\renewcommand{\baselinestretch}{1.25}
\selectfont

\begin{center}

{\Large\textbf{Abstract}}

\end{center}

\raggedright



\begin{textblock}
\item
  The common phrase `representation matters' 
  asserts that 
  media has a measurable and important
  impact on civic society's perception of self and others.
\item
  The representation of health in media,
  in particular,
  may reflect and
  perpetuate a society's disease burden.
\item 
  Here, for the top 10 major causes of death in the United States,
  we examine how cinematic representation of
  overall and by-gender mortality diverges 
  from reality.
\item
  Using crowd-sourced data on film deaths from 
  Cinemorgue Wiki,
  we employ natural language processing (NLP) techniques to analyze
  shifts in representation of deaths in movies versus the 2021 National
  Vital Statistic Survey (NVSS) top ten mortality causes.
\item 
  Overall, movies strongly overrepresent suicide and, to a lesser degree, accidents.
\item 
  In terms of gender,
  movies overrepresent men and underrepresent women for nearly every major mortality cause, including heart disease and cerebrovascular disease. 
 \item 
 The two exceptions for which women 
 are overrepresented are suicide and accidents.
\item
  We discuss the implications of under- and over-representing
  causes of death overall and by gender, as well as areas of future research.
\end{textblock}

\renewcommand{\baselinestretch}{1}
\selectfont

\twocolumn

\restoregeometry

\clearpage

\section{Introduction}

Cardiovascular disease is the number one cause of death in both men
and women in the United States according to the National Survey of
Vital Statistics (NVSS), published annually by the Centers for Disease
Control~\cite{curtin_deaths_2024}.
However, men and women can experience different risk factors related
to cardiovascular disease, as well as present different symptoms when
experiencing a cardiovascular event~\cite{isakadze_addressing_2019}.
In the largest study to date of training physicians, more than 70\%
report that they are \textit{not} explicitly taught about gender
differences across disease presentation~\cite{dhawan_sex_2016}.
Furthermore, in a study that provided physicians with gender-blinded
lists of risk factors specifically for cardiovascular disease, most
physicians systematically ``under-risked'' female patients for heart
disease~\cite{mosca_national_2005}.
In recent years, the American Heart Association has highlighted the
importance of understanding these differences with calls for an
overhaul in how physicians consider gender in diagnosis of
cardiovascular disease~\cite{wenger_call_2022}.
Furthermore, physicians in the United States and elsewhere face
increasing patient loads, and often employ formal and informal
heuristics to speed up treatment or facilitate easier patient
hand-off between specialities.
While heuristics borne of substantive modeling and research are
generally considered positive for patient outcomes~\cite{marewski_heuristic_2012},
unconscious or conscious bias held by
healthcare providers can create availability heuristics that lead to
worse patient outcomes.
For example, in a widely cited study, false beliefs of racial
differences between black and white patients by white medical
students and residents predicted lower accuracy pain treatment for
black patients~\cite{hoffman_racial_2016}.
Outside of a formal healthcare setting, how health and disease are
understood by the population-at-large can be readily influenced by
numerous factors, including media portrayals.

Representation in media is central to how we perceive our society,
including health.
Media portrayal and representation influences thoughts, behaviors,
and actions towards certain groups, ideas, and phenomena.
That is, the ``[m]edia, in short, [is] central to what ultimately
come to represent our social realities''~\cite{dow_gender_2006}.
This connection has been studied across many demographic categories
and aspects of social reality. 
Recent research has examined how acute myocardial infarction (a form of heart attack) was portrayed in 100 popular films, demonstrating the general inaccuracy of its depiction of symptoms as well as its depiction by gender~\cite{shaw_portrayal_2024}. 
Aside from that study, in terms of gender demographics, most research tends to focus on negative
representation of woman in media, specifically in terms of sexual
objectification; ``very few studies appear to be available on the
relationship between media representation and non-sexual
objectification" of women~\cite{santoniccolo_gender_2023}.
Similarly, many aspects of social reality, like economic, health, and
well-being outcomes, have been studied with regards to media
representation (though not explicitly through the lens of gender).
For example, there is a rich history of research on how
representation in media influences the commission of intentional
self-harm~\cite{sisask_media_2012}.
Summarily, media typically does not accurately portray official
suicide data and the most comprehensive systematic review on the
topic connects media reporting and suicidality, with current evidence
suggesting increased suicidality trends with increased reporting of
suicide~\cite{sisask_media_2012}.

Aside from suicidality, there is a diversity of studies on health
issues and their media depiction; a non-exhaustive survey of the
literature includes a study on the depiction of obesity in media and
mortality in indigenous Australians~\cite{islam_indigenous_2016}, a
study of how bisexual individuals' mental health is impacted by
depictions of bisexual people in media
\cite{johnson_bisexuality_2016}, and globally, a scoping review on how
media representation of major corporations influence non-communicable
disease risk assessment and public health policy making related to
consumer products~\cite{weishaar_why_2016}.
Thus, while there is ample research on health representation in
various forms of media, and research on gender representation in
media~\cite{bamman2024measuring,bleakley2012trends,anderson2016largest,smith2014gender,smith2015inequality,lindner2015million}, there is little research that explores the overlap between
health outcomes and gender representation in media, a gap we aimed
to address in this study.

To examine the connection between media and mortality in this study,
we downloaded counts by year of all-time cinematic deaths (referred to
here as film deaths) from the crowd-sourced Cinemorgue Wiki database
\cite{noauthor_cinemorgue_2024} and compared them to the 2021 National
Vital Statistics Survey (NVSS)~\cite{curtin_deaths_2024} counts of top
ten causes of mortality by gender.
Using natural language processing (NLP), we extracted the Cinemorgue
counts, classified the deaths by gender, and consolidated this data
against the NVSS counts.
We then analyzed the comparative ratios by gender between real-life
and film are and considered significant differences in these ratios
using the Chi-Square Test.

\section{Methods}

There are two main data sources for this project: the Cinemorgue Wiki
database of film deaths~\cite{noauthor_cinemorgue_2024} and the 2021
NVSS Leading Causes of Mortality Report~\cite{curtin_deaths_2024}.
 
\subsection{Collecting and Cleaning Data}


First, we identified all-time films deaths using the Cinemorgue Wiki
\cite{noauthor_cinemorgue_2024} which is a crowd-sourced database that
allows any members of Fandom to update death by cause by film, with a
short description of the death.
The site is organized by death-type or by actor (not by character) and
each entry is a short 1-2 sentence description of the film death.
It is inclusive of \textit{any} movies publicly entered, and thus not
limited to American film deaths only.
Conveniently, Cinemorgue has historically provided an .xml download
of their entire website (current as of August 16, 2023) and other
researchers have previously classified cinematic deaths using FBI
homicide specifications~\cite{cinemorgue_parsing}.

Using this existing script, we adjusted the classifying portion to
identify deaths according to the Centers for Disease Control National
Vital Statistics System's top mortality causes by year for
2021~\cite{curtin_deaths_2024}.
Our script parsed the description of each film and reduced it to word
stems by film entry.
We used generally broad word stems that captured the colloquial terms
describing different mortality causes (for example ``heart attack''
AND ``heart failure'').
We then categorized film deaths by gender using a reasonably
comprehensive and inclusive classifier from PyPi project
\texttt{gender-guesser}, which was trained on a representative
international database of names that was independently validated by
native speakers of many
languages~\cite{noauthor_gender-verification_nodate}.
When applying \texttt{gender-guesser} in this study, all 40,000 plus
names were included to account for various
languages and ethnic backgrounds of actors represented in film (the
Cinemorgue Wiki includes movies from many different countries).
Furthermore, the code allowed for the classification of mostly male,
mostly female, androgynous and unknown names.
Table~\ref{table:count_before} enumerates the initial name
classification.

\begin{table}[t] 
\begin{center}

\begin{tabular}{l|c|c}
\bf{Gender} & \bf{Count Before} &\bf{Count After} \\
\hline
Female & 15,673 & 17,451\\
Male & 39,627 & 42,458 \\
Unknown & 7,383 & --- \\
Mostly Male & 2,831 & --- \\
Mostly Female & 1,778 & ---\\
Androgynous & 1,243 & ---  \\
Non-male/Non-female & --- & 8,626 \\
\end{tabular}
\caption{Results of the gender classifier, \texttt{gender-guesser}, before and after consolidation.} \label{table:count_before}
\end{center}
\end{table}

Based on a random manual review of classification, ``mostly male'' and
``mostly female'' seemed to accurately categorize well-known actors,
``mostly male'' was grouped with male and ``mostly female'' was
grouped with female.
Based on the size of the data and this classifier, we removed
androgynous or unknown names (collectively referred to here as
``other''), discussed in limitations.


The 2021 top ten leading causes of death in the United States are
depicted in Table~\ref{table:2020_death}~\cite{curtin_deaths_2024}.
This list actually comprises 13 leading causes of death because the
9th and 10th leading causes of death by gender differ, and a category
of ``other'' is including to reflect the remaining causes.
If absent from the official report, this data was queried directly
from the CDC WONDER database~\cite{wonder} using the same ICD-10
(International Classification of Disease) parameters from the
NVSS~\cite{curtin_deaths_2024}.
We calculated the ratio of mortality by gender by performing a simple
ratio of male or female deaths by mortality cause over total deaths
by that cause.
It is important to note that the top 10 leading mortality causes
differ year-to-year, and 2021 was the second year to include
COVID-19 deaths.
Additionally, the medical terms describing these causes of death are
extensive and often highly technical; for ease of understanding, we
provide additional notations to specify the colloquial terms for
these diseases in italics.

\begin{table*}[t]

\begin{center}

  \begin{tabular}{
      p{0.4\textwidth}
      |
      >{\raggedleft\arraybackslash}p{0.1\textwidth}
      >{\raggedleft\arraybackslash}p{0.1\textwidth}
      >{\raggedleft\arraybackslash}p{0.09\textwidth}
      >{\raggedleft\arraybackslash}p{0.09\textwidth}
      >{\raggedleft\arraybackslash}p{0.09\textwidth}
    }
    \textbf{Mortality Cause}\newline
    \textcolor{grey}{Overall rank. Type (Female rank, Male rank)}
    &
    \textbf{Total}
    &
    \textbf{Female Count}
    &
    \textbf{Female~\%}
    &
    \textbf{Male Count}
    &
    \textbf{Male~\%}
    \\
    \hline
    1. Diseases of the Heart
    {(1,1)}
    &
    695,547
    &
    310,661
    &
    44.6\% 
    &
    384,886
    &
    55.4\%
    \\
    \hline
    2. Malignant Neoplasm \textit{(Cancer)}
   {(2,2)}
    &
    605,213
    &
    286,543
    &
    47.3\%
    &
    318,670
    &
    52.7\%
    \\
    \hline
    3. COVID-19 {(3,3)} 
    & 
    416,893
    & 
    180,283 
    & 
    43.2\% 
    &  
    236,610 
    &  
    56.8\%\\
    \hline
    4. Accidents (unintentional injuries) {(6,4)} 
    & 
    224,935 
    & 
    75,333 
    & 
    33.5\% 
    & 
    149,602 
    & 
    66.5\%  \\
    \hline
    5. Cerebrovascular Disease \textit{(stroke, brain hemorrhage)} {(4,5)} 
    & 
    162,890 
    & 
    92,038 
    & 
    56.5\% 
    & 
    70,852 
    & 
    43.5\% \\
    \hline
    6. Chronic Lower Respiratory Disease {\textit{(COPD, emphysema, etc.)}} {(7,6)} 
    & 
    142,342 
    & 
    74,814 
    & 
    52.6\% 
    & 
    67,528 
    & 
    47.4\% \\
    \hline
    7. Alzheimer's Disease {(5,8)} 
    & 
    119,399 
    & 
    82,424 
    & 
    69.0\% 
    & 
    36,975 
    & 
    31.0\% \\
    \hline
    8. Diabetes Mellitus {(8,7)} 
    & 
    103,294
    & 
    44,666 
    & 
    43.2\% 
    & 
    56,628 
    & 
    54.8\% \\ 
    \hline
    9. Chronic Liver Disease and Cirrhosis {(11,9)} 
    & 
    56,585 
    & 
    20,878 
    & 
    36.9\% 
    & 
    35,707 
    & 
    63.1\% \\
    \hline
    10. Nephritis {(9,10)} & 54,358 & 25,769 & 47.4\% & 28,589 & 52.6\% \\
    \hline
    Suicide (intentional self-harm) {(15,8)} & 48,183 & 9,825 & 20.4\% & 38,358 & 79.6\% \\
    \hline
    Essential hypertension {(10,15)} & 42,816 & 22,730 & 53.1\%  & 20,086 & 46.9\% \\
    \hline
    Other & 791,776 & 400,159 & 50.5\% & 391,617 & 49.5\% 
  \end{tabular}

\end{center}

\caption{
  Percentage by Gender of the National Vital Statistics System 2021 Leading Causes of Mortality in the United States.}

\label{table:2020_death}

\end{table*}

\subsection{Analysis of Gender Ratios: Visualization and Chi-Square Testing}

In both data sets, we calculated the ratio of mortality by gender by cause  by performing a simple ratio of male death (or female death) by mortality cause over total deaths by that cause. Thus, the sum of all ratios should equal 1. From the ratios of mortality by gender in film and NVSS data, we constructed a bar graph visualization. Then, we explored the significance of difference in gender ratio by cause of death between film and real-life using the Pearson's Chi-Square Test~\cite{franke_chi-square_2012}\cite{mchugh_chi-square_2013}.  

It is important to underscore \textit{how} to interpret significant
$p$-values in this
context~\cite{wasserstein_asa_2016,wasserstein_moving_2019,greenland_statistical_2016}.
Accordingly, $p$-values $< 0.5$ do not ``measure the probability that
the studied hypothesis is true, or the probability that the data were
produced by random chance alone'', and are merely ``one approach to
summarizing the incompatibility between a particular set of data and
a proposed model for the data''~\cite{wasserstein_asa_2016}.
That is, if there is a significant $p$-value for the Chi-Square test
for any given cause of mortality, this means that the data seems to
suggest there is a significant difference between expected and
observed frequencies of causes of death by gender between film and
real-life but does not establish causality.

\section{Results}
\begin{figure*}
    \centering
    \includegraphics[width=\textwidth]{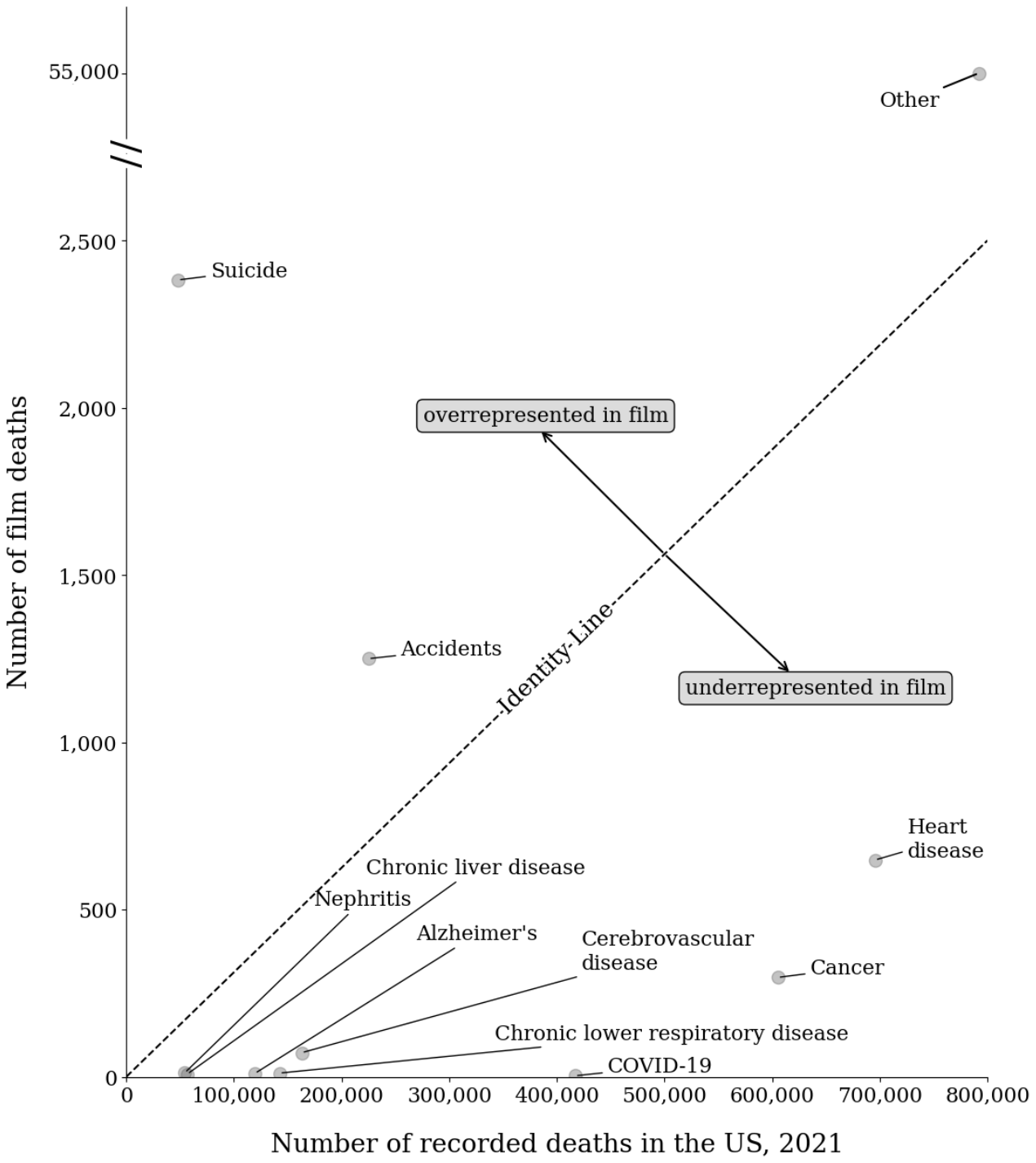}
    \caption{
      Overall number of total deaths in Cinemorgue versus National
      Vital Statistics Survey 2021 by Cause of Mortality. Of total film deaths, suicide and accidents are over-represented; all other causes of death are underrepresented in film.
    }
    \label{fig:deathstotal}
\end{figure*}

In general, the Cinemorgue dataset under-represents most real-life
leading mortality causes as compared to cinematic mortality causes by
proportion of deaths in the database.
 There are two exceptions where suicide and accidents are
 over-represented.
 Figure~\ref{fig:deathstotal} depicts the identity line or line of no
 difference at which representation would be equal between both
 sources.

\begin{figure}[htbp] 
  \centering
  \includegraphics[width=\columnwidth]{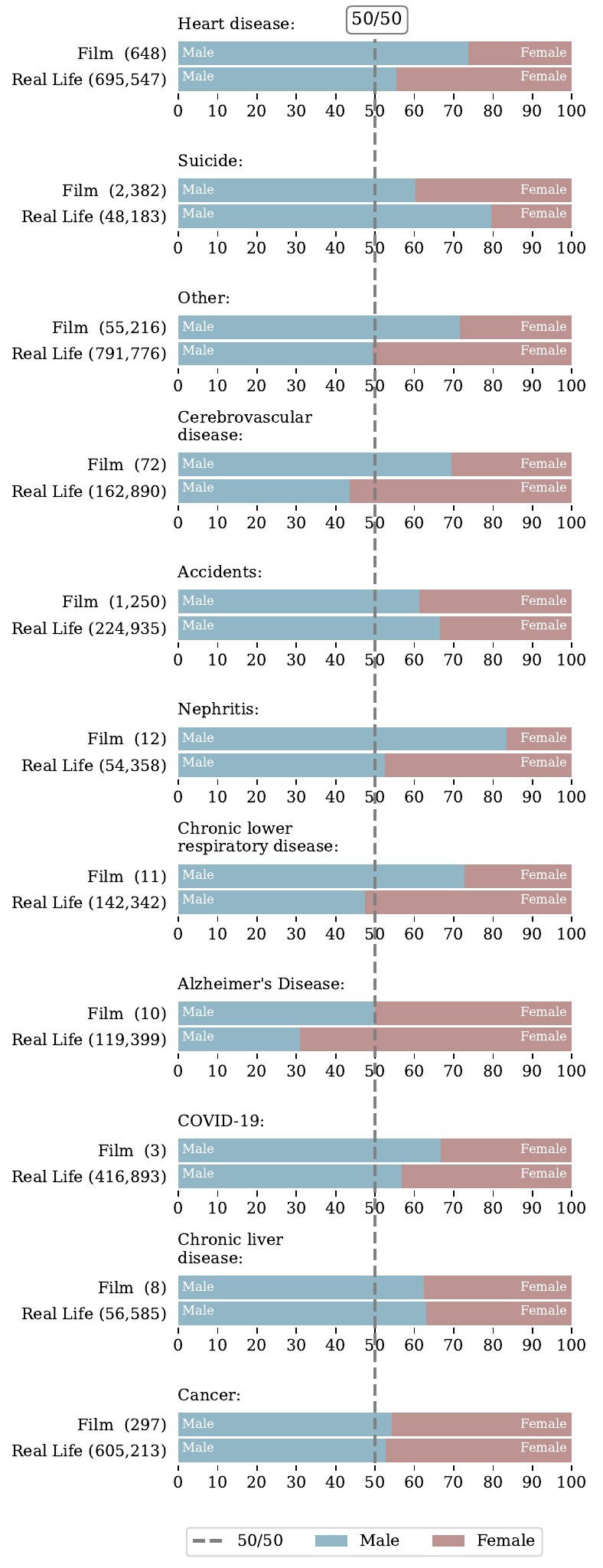}
  \caption{
    Top Ten Mortality Causes by Gender in Film versus Real-Life: 
    Female suicide deaths and male accidents, cerebrovascular disease, and heart disease are over-represented in film. Men are over-represented in all other causes of death. 
    We only include causes of death present in Cinemorgue dataset.
  }
  \label{figure:mortalitybygender}
\end{figure}

When considering gender in addition to raw count by database, the
differences between film and real-life are also apparent.
Figure~\ref{figure:mortalitybygender}: Top Ten Mortality Causes by
Gender in Film versus Real-Life'' is a sub-plot panel depicting the
relative proportions of 2021 NVSS leading causes of mortality by
gender versus all-time film deaths by gender from Cinemorgue.
Across almost all causes of death except cancer and chronic liver
disease, there are visible and apparent shifts between film
representation and real-life representation, with women
under-represented in film across nearly all categories except for
suicide.
Diabetes and essential hypertension are not represented in the
cleaned Cinemorgue dataset.
Note: the counts used to construct the ratio shift figure are
contained in Table~\ref{table:cinemorgue_death}: Percentage and
Count for Leading Causes of Mortality in the United States in
Cinemorgue Wiki Database'' in the appendix.

In Table~\ref{table:chisquare},
we show the significance (if observed frequency
 $> 5$) of the ratio shift for each cause of death.
 The ratios by gender of accidents, cerebrovascular disease, diseases
 of the heart, suicide (intentional self-harm), and the combined
 category of all other causes of mortality for film deviate
 significantly from that in real-life.
 As previously mentioned, men are over-represented in film as compared
 to women for each of the aforementioned causes of mortality, barring
 suicide.

\begin{table}[t] 

  \begin{tabular}{
      p{0.50\columnwidth}
      |
      p{0.15\columnwidth}
      |
      p{0.25\columnwidth}
    }
    \textbf{Cause of Death} & \textbf{$\chi^2$}  & \textbf{$p$-value}\\
    \hline
    Accidents*& 1.6$\times$$10^{1}$ & 7.0$\times$$10^{-5}$ \\
    \hline
    Alzheimer's Disease & NA & NA \\
    \hline
    COVID-19 & NA & NA \\
    \hline
    Cancer & 2.8$\times$$10^{-1}$ & 0.59 \\
    \hline 
    Cerebrovascular disease* & 2.0$\times$$10^{1}$ & 8.9$\times$$10^{-6}$\\
    \hline
    Chronic liver disease & NA & NA \\ 
    \hline
    Chronic lower respiratory diseases & NA & NA \\
    \hline
    Diseases of heart* & 87.6 & $<$$10^{-20}$ \\
    \hline
    Nephritis & NA & NA \\
    \hline
    Other* & 1.1$\times$$10^{4}$& $<$$10^{-20}$ \\
    \hline
    Suicide* & 5.5$\times$$10^{3}$ & $<$$10^{-20}$ \\
  \end{tabular}

  \caption{
    Chi-Square Test: Statistic \& $p$-values.
    NA applies to cause of death categories in which there
    were less than 5 observations in films, and thus the
    Chi-square test was not appropriate~\cite{franke_chi-square_2012}. Significant findings (where film deaths by gender are not representative of real life) are annotated with 
    *.
  }
  \label{table:chisquare}    
\end{table}

\section{Discussion}

Film and real-life mortality differ significantly for some, but not
all, leading causes of death.
As expected, film is not representative of real-life even setting aside
gender considerations.
From a practical perspective, this makes sense in the context that
high-tempo, salacious, or dramatic deaths are likely far more
interesting to an audience than everyday causes of mortality.
In film, for example, the proportion of ``other'' deaths represented
far outpaces the actual leading cause of death (diseases of the
heart) at a factor of over 800.
When considering gender, there are significant differences in ratio
of represented death for accidents, cerebrovascular disease, diseases
of the heart, suicide (intentional self-harm) and all other deaths
not captured using the top ten leading causes of death classifier,
among sufficiently large samples within the Cinemorgue Wiki database.

While our results do not indicate causality, they provide insight to
the representation of mortality in popular media.
From visualizations and statistical comparison of ratios, there is
meaningful descriptive power in the ratios of film versus real-life
representation of mortality by gender.
In view of recent efforts by the American Heart Association to
provide better equality of cardiovascular care for
women~\cite{wenger_call_2022} as well as general research regarding
treatment differences by gender~\cite{dhawan_sex_2016}, the
under-representation of film deaths by gender demonstrate that death
is also lacking in equitable representation on screen.
As there is general consensus in the literature that stereotypes can
be exceedingly harmful and contribute to sexism, harassment, and
violence against women~\cite{santoniccolo_gender_2023} as well as
other gender identities~\cite{hughto_negative_2021}, it is reasonable
to postulate that lack in representation in film \textit{could}
exacerbate the culture of difference regarding disease treatment by
gender.

By cause of mortality, some interesting patterns emerge.
For the category of ``other'', men are over-represented in the
Cinemorgue Wiki, which is consistent with their over-representation
writ large in the Wiki (regardless of cause of death).
This could indicate that men are represented more often on screen
generally, and represented more often in violent deaths, as
previously discussed.
For diseases of the heart, men are seemingly over-represented in
film, which agrees with the American Heart Association's recent
findings on an under-representation and lack of understanding of
women experiencing heart disease~\cite{wenger_call_2022}. Likewise, men are over-represented for death by cerebrovascular disease. 
However, suicide and accidents are over-represented for women in film.
Based on the suggestibility and risk associated with suicide
representations~\cite{sisask_media_2012}, this pattern indicates that
future follow-up is needed to understand the relationship
between representation and acts of suicide in women.

\subsection{Limitations}

Limitations of this study center on classification and cleaning of the
Cinemorgue film death dataset, as well as the data itself.
First, we utilized only film deaths in this study, on the assumption
that films generally have wider ranges of audiences (not confined to
certain streaming platforms) and have a similar format.
However, the Cinemorgue Wiki database includes all manners of
video-represented deaths, including television and video games, which
were excluded in this study based on the assumption that video game
deaths may over-represent homicides, due to the often violent nature
of many video games.
Furthermore, we compare the Cinemorgue Wiki database all-time counts
to single point in time (2021 NVSS) mortality counts; considering
that movies are not only watched in the year in which they are
filmed, we consider this a reasonable assumption even though movie
popularity changes over time.

Additionally, our removal of androgynous names may have skewed the
resultant data; as shown previously in Table~\ref{table:count_before}, 13\%
of data is effectively ``thrown out'' of the analysis because it was
not classifiable as ``female'' or ``mostly female'' or ``male'' or
``mostly male''.
While using binary categories is consistent with the CDC's current
reporting causes by mortality, reducing gender to solely male and
female categories does not capture the diversity of gender
identities. Considering the connection between media representation and suicide, as well as the significantly higher instance of suicidal ideation in transgender and non-binary youth than cisgender youth~\cite{suarez_disparities_2024}, indicates that further research beyond this study is important for understanding representation of health outcomes for transgender and nonbinary individuals. 
In terms of classifying cause of mortality in the Cinemorgue Wiki
database, the search terms used on the stemmed text were specific but
very broad in order to prevent an under-count of deaths, which could
have skewed resultant counts. Intrinsically, the Cinemorgue Wiki database is also subject to the
normal risks of crowd-sourced data.
Members of the public that are heterogeneous in knowledge, skill, and
geographical location comprise the
crowd~\cite{lenart-gansiniec_understanding_2023} that generates the
entries in the Cinemorgue Wiki; biases in any of these
characteristics could contribute to a geographically skewed or low
fidelity dataset.
While CDC data from the United States was used to calculate gender
proportions, every film death in the database was included, even if
the film was not English-language.

\subsection{Future Work}

Future research should expand the corpus of examined work and include
mortality representation across all forms of media, including books,
songs and social media.
This would provide a more global perspective on the impact of
media---writ large---and capture populations that may not interact
with cinematic media.
With enough data, the Granger causality test---used to forecast
time-series from other
time-series~\cite{shojaie_granger_2022}---could be employed to
understand whether media representation of death by gender lags or
leads actual causes of mortality by gender.
This would contribute to a better understanding of the interplay
between media representation and health outcomes over time.
Relatedly, visualizations like a Sankey flow diagram---which provides
visualization of time-series-type data in lieu of a formal time
series analysis~\cite{otto_overview_2022}---could provide further
opportunities for interpretation of how gender representation has
morphed over time.
Randomized controlled trials centering on these ideas (perhaps
priming physicians with biased images prior to diagnosis of case
notes) could provide support for the case of causality between
representation and medical diagnosis and treatment.

\section{Conclusion}

Our study compared mortality causes by gender in film and real-life
using Cinemorgue Wiki data on all-time film deaths and NVSS mortality
by cause data from 2021 in the United States.
Based on comparison of gender ratios by cause of death, we identified
significant differences in representation for accidents,
cerebrovascular disease, diseases of the heart, and suicide
(intentional self-harm).
Considering current struggles in appropriately managing diagnosis and
treatment of disease by gender, highlighting how representation of
mortality is skewed by gender in film may encourage a more equitable
culture surrounding gender and health.
Future work should expand this query to include other forms of media
to understand the interplay of representation and mortality.



\section*{Acknowledgments}

The authors are grateful for support from the
National Science Foundation (Award \#2242829)
and the Massachusetts Mutual Life Insurance Company.

\bibliography{\filenamebase}


\onecolumn

\appendix
\section{Appendices}

\setcounter{page}{1}
\renewcommand{\thepage}{A\arabic{page}}
\renewcommand{\thefigure}{A\arabic{figure}}
\renewcommand{\thetable}{A\arabic{table}}
\setcounter{figure}{0}
\setcounter{table}{0}

\renewcommand{\thesection}{A\arabic{section}}
\setcounter{section}{0}

\section*{Appendix}

\begin{table}[h]

\begin{tabular}{p{0.4\textwidth}|p{0.07\textwidth}p{0.1\textwidth}p{0.1\textwidth}p{0.09\textwidth}p{0.1\textwidth}}

\bf{Mortality Causes} & \bf{Total} & \bf{Female \%} & \bf{Female Count} & \bf{Male \%} & \bf{Male Count} \\
\hline
1. Diseases of the Heart & 648 & 26.4\% & 171 & 73.6\% & 477 \\
\hline
2. Malignant Neoplasm \textit{(Cancer)} & 297  & 45.8\% & 136 & 54.2\% & 161  \\
\hline
3. COVID-19 & 3 & 33.3\% & 1  & 66.7\% & 2 \\
\hline
4. Accidents (unintentional injuries)& 1,250 & 38.8\% & 485 & 61.2\% & 765 \\
\hline
5. Cerebrovascular Disease \textit{(stroke, brain hemorrhage)} & 72 & 30.6\% & 22 & 69.4\% & 50\\
\hline
6. Chronic Lower Respiratory Disease {\textit{(COPD, emphysema, etc.)}} & 11 & 27.3\% & 3 & 72.7\% & 8  \\
\hline
7. Alzheimer Disease & 10 & 50\% & 5 & 50\% & 5 \\
\hline
8. Diabetes Mellitus & --- & --- & --- & --- & --- \\ 
\hline
9. Chronic Liver Disease & 8 & 37.5\% & 3 & 62.5\% & 5\\
\hline
10. $\ddagger$Nephritis & 12 & 16.7\% & 2 & 83.3\% & 10\\
\hline
$\dagger$Suicide (intentional self-harm) & 2,382 & 39.7\% & 945 & 60.3\% & 1437\\
\hline
$\ddagger$Essential hypertension & --- &  --- & --- & --- & ---\\
\hline
\^{}Other & 55,216 & 28.4\% & 15,678 & 71.6\% & 39,538
\end{tabular}
\caption{Percentage and Count for Leading Causes of Mortality in the United States in Cinemorgue Wiki Database}\label{table:cinemorgue_death}

    {\raggedright \small{$\dagger$Denotes \textit{top 10} cause of death for males only}} 

    {\raggedright \small{$\ddagger$Denotes \textit{top 10} cause of death for females only}} 

    {\raggedright \small{\^{}``Other'' causes of death are all deaths not captured in the categories above}}

\end{table}

\begin{table*}[h]

\begin{tabular}{p{0.4\textwidth}|p{0.17\textwidth}|p{0.10\textwidth}}

\bf{Mortality Causes} & \bf{Cinemorgue}  & \bf{NVSS}\\
\hline
1. Diseases of the Heart & 1\% & 19.6\%  \\
\hline
2. Malignant Neoplasm \textit{(Cancer)} & $<$1\% & 17.1\% \\
\hline
3. COVID---19 & $<$1\% & 11.8\% \\
\hline
4. Accidents (unintentional injuries) & 2.1\% & 6.4\% \\
\hline
5. Cerebrovascular Disease \textit{(stroke, brain hemorrhage)} & $<$1\% & 4.6\%\\
\hline
6. Chronic Lower Respiratory Disease {\textit{(COPD, emphysema, etc.)}} & $<$1\% & 4.0\%  \\
\hline
7. Alzheimer Disease & $<$1\% & 3.4\% \\
\hline
8. Diabetes Mellitus & --- & 2.9\% \\ 
\hline
9. Chronic Liver Disease & $<$1\% & 1.6\%\\
\hline
10. $\ddagger$Nephritis & --- & 1.5\% \\
\hline
$\dagger$Suicide (intentional self---harm) & 3.9\% & 1.4\%\\
\hline
$\ddagger$Essential hypertension & --- & 1.21\% \\
\hline
\^{}Other & 92.4\% & 24.9\%
\end{tabular}
\caption{Percentage of Leading Causes of Mortality in the United States in Cinemorgue Wiki Database versus NVSS}\label{table:cinemorgue_death}

    {\raggedright \small{$\dagger$Denotes \textit{top 10} cause of death for males only}} 

    {\raggedright \small{$\ddagger$Denotes \textit{top 10} cause of death for females only}} 

    {\raggedright \small{\^{}``Other'' causes of death are all deaths not captured in the categories above}}

\end{table*}

\end{document}